\documentclass[12pt]{article}
\hoffset = - 0.5 in
\begin{document}
\title{
 DYNAMO IN HELICAL MHD TURBULENCE: Quantum Field Theory Approach}

\author{M. Hnatic$^{1}$, M. Jurcisin$^{1,2}$, M. Stehlik$^{1}$ \\
$^{1}$ {\small{\it  Institute of Experimental Physics of SAS,
 Watsonova 47, 043 53, Ko\v{s}ice, Slovakia}}, \\
$^{2}$ {\small{\it  BLTP, Joint Institute for Nuclear Research, 141
980 Dubna, Russian Federation}} }
\date{}

\maketitle
\begin{abstract}
A quantum field model of helical MHD stochastically forced by
gaussian hydrodynamic, magnetic and mixed noices is investigated.
These helical noises lead to an exponential increase of magnetic
fluctuations in the large scale range. Instabilities, which are
produced in this process, are eliminated by spontaneous symmetry
breaking mechanism accompanied by creation of the homogeneous
stationary magnetic field.
\end{abstract}

\section{Introduction}

Quantum field theory method including renormalization group (RG)
approach has been successfully used for the theoretical explanation
of various phenomena in developed turbulence (see \cite{vas} and
wherein references).

In this paper  the quantum field model of helical MHD is
investigated. As a starting point we consider the Navier-Stokes
equation for the velocity field and the equation for magnetic field
which are driven by gaussian random forces with a given $2\times2$
matrix $D$ of the hydrodynamic, magnetic and mixed noise
correlators, respectively.

In ref. \cite{MHD} (see also \cite{four}) the multiplicative
renormalizability of the quantum field model of non-helical MHD
turbulence has been proved and the RG approach has been applied to
study the asymptotic behavior of the model considered. The existence
of two infrared-stable fixed points has been established. These
points govern the two critical regimes: the magnetic regime and the
kinetic one (the later being of the Kolmogorov type).

The critical properties of the helical MHD are not known in the case
of an arbitrary noise matrix $D$. To provide the multiplicative
renormalizability and consequent application of RG it is necessary
to extend the theory adding the extra dissipative terms with new
helical Prandtl numbers \cite{RG91}. Therefore, also a critical
behavior of the helical MHD is more complicated. A priori, the
existence of the former stable regime of the Kolmogorov type is not
clear. In the following the existence of the  critical regimes
mentioned above for ordinary MHD is demonstrated for the helical
one.

There is an additional problem in the helical MHD: the instability
of the theory which is induced by the exponential increase of the
magnetic fluctuations in the large scales range (see \cite{tur}, for
example). The elimination of this instability leads to formation of
a large-scale magnetic field known as the turbulent dynamo. Removal
of the instability in quantum field formulation of helical MHD can
be achieved by means of a nice and very well known spontaneous
symmetry breaking mechanism followed by the creation of homogeneous
stationary magnetic field. The special case, when only the
hydrodynamic noise does not vanish, was analyzed in  \cite{dyn}.

In this paper the value of spontaneous magnetic field ${\bf c}$ in
one loop approximation for matrix $D$ of noises in generic form.
This value has been found from the conditions of overall exponential
instabilities elimination in the steady state.

The dynamo mechanism is accompanied by the appearance of an "exotic"
term in the linearized equation for magnetic field. This term causes
the linear growth in time $t$ of the amplitudes of Alfv\'en waves
for small wave vectors ${\bf k}$ in the direction orthogonal to the
plane of ${\bf k}$ and ${\bf c}$.  Due to the viscosity terms this
growth is transformed into long-lived pulses of the type $t \exp(-i
\beta t)\exp (-\alpha t)$ with small $\alpha>0$ and $\beta.$

\section{The formulation of the problem}

The interaction of electrically neutral conductive turbulent
incompressible fluid with the magnetic field with the magnetic field
is described by the stochastically forced MHD equations \cite{MHD}:
\begin{eqnarray}
\nonumber \partial_t {\bf v} &=& \nu \Delta {\bf v} -({\bf v}
\partial){\bf v} + ({\bf b}\partial){\bf b}-
\partial p+ {\bf F^{v}}\\
\partial_t {\bf b} & = & \nu'\Delta{\bf b}-({\bf v}
\partial){\bf b} + ({\bf b}\partial){\bf v}+
{\bf F^{b}}\, , \label{ns}
\end{eqnarray}
together with the incompressibility conditions
\begin{equation}
\nabla \cdot {\bf v}=0,\,\,\,\, \nabla \cdot {\bf F^{v}}=0,\,\,\,\,
\nabla \cdot {\bf F^{b}}=0. \label{incom}
\end{equation}
The first equation is the well-known Navier-Stokes equation for the
divergence free velocity field ${{\bf v}}(x)=\{v_i({\bf x},t)\}$
with the additional nonlinear contribution of the Lorentz force (p
is a sum of both hydrodynamic and magnetic pressure per unit mass).
The second equation for magnetic field ${\bf b}(x)= \{b_i({\bf
x},t)\}$ (it is connected with magnetic induction $ {\bf B}$ by the
relation $ {\bf b}={\bf B}/\sqrt{4\pi\varrho}$, where $\varrho$ is
the fluid density and $\mu$ is its magnetic permeability) follows
from the Maxwell equations for continuous medium. The magnetic
diffusion coefficient $\nu'$ is connected with the coefficient of
molecular viscosity by relation $\nu'=u\nu$ with dimensionless
magnetic Prandtl number (PN) $u^{-1}$.

External random forces ${\bf F^{v}}, {\bf F^{b}}$ are assumed to
have a Gaussian distribution with $<F> =0$ and they are defined by
$2\times 2$ matrix of the noise correlators $D = <F F>$. The matrix
elements are: the hydrodynamic  $ D^{{\bf v}{\bf v}}$ noise, the
magnetic $D^{{\bf b}{\bf b}}$ one and the mixed $D^{{\bf v}{\bf b}}$
one.

The problem (\ref{ns}) is equivalent to  a quantum theory with the
doubled number of the fields $\Phi=\{{\bf v},{\bf b},{\bf v}',{\bf
b}'\}$ and the action functional \cite{MSR1,MSR}:
\begin{eqnarray}
\nonumber S(\Phi) & = & \frac{{\bf v}'D^{{\bf v}{\bf v}}{\bf
v}'}{2}+ \frac{{\bf b}'D^{{\bf b}{\bf b}}{\bf b}'}{2}+ \frac{{\bf
v}'D^{{\bf v}{\bf b}}{\bf b}'}{2}+ \frac{{\bf b}'D^{{\bf b}{\bf
v}}{\bf v}'}{2}+
{\bf v}' [-\partial_t{\bf v}+\nu_0\Delta{\bf v} \\
& & -({\bf v}\partial){\bf v}+({\bf b}\partial){\bf b}]+ {\bf b}'
[-\partial_t{\bf b}+u_0\nu_0\Delta{\bf b}- ({\bf v}\partial){\bf
b}+({\bf b}\partial){\bf v}]\, , \label{zd}
\end{eqnarray}
where ${\bf v}', {\bf b}'$ are some auxiliary vector fields.
Hereafter in the similar expressions, the integration over ${\bf
x}$, $ t$ and the traces over the vector indices are implied. As it
is usual in QFT, the action (\ref{zd}) is considered to be
unrenormalized with the bare constants marked by the subscript
$"0".$ The basic objects of the study are the Green functions of the
fields $\Phi$ (the correlation functions and response functions in
the terminology of the original problem (\ref{ns})). They can be
determined as functional derivatives with respect to an external
sources $ A=\{A^{\bf v}, A^{\bf b}, A^{{\bf v}'}, A^{{\bf b}'}\}$ of
the generating functional $G(A) = \int D\Phi \exp[S(\Phi)+A\Phi]$,
i.e., they are the functional averaged values of the corresponding
number of the fields $\phi$ with a weight $ \exp[S(\phi)]$. Here, $
D\Phi $ denotes the functional measure of the integration over the
fields $\Phi$ with all normalization coefficients.

We have to choose a concrete form of  $D$ in the wave
vector-frequency $({\bf k},\omega)$ representation. The noises are
transversal for the incompressible fluid. The action for helical MHD
can possess scalar terms as well as pseudoscalar ones. Hence, the
tensor structure of all noises is a linear combination of both
tensor and pseudotensor. Then the correlators have the form:
\begin{eqnarray}
D_{js}^{{\bf v}{\bf v}}&=&g_0\nu_0^3k^{1-2\epsilon}{\bf
P}^1_{js}\,, \nonumber \\
D_{js}^{{\bf b}{\bf b}}&=&g'_0\nu_0^3k^{1-2a\epsilon}{\bf
P}^2_{js}\,,
\nonumber \\
D_{js}^{{\bf b}{\bf v}}&=& D_{js}^{{\bf v}{\bf b}}=
g''_0\nu_0^3k^{1-(1+a)\epsilon}{\bf P}^3_{js}. \label{ks}
\end{eqnarray}
Here, ${\bf P}^r_{js}=P_{is}+i\rho_r\varepsilon_{jsl}k_l/k$, where $
P_{is}=\delta_{is}-k_i k_s/k$ stands for transverse projector and $
\varepsilon_{isl}$ is Levi-Civita pseudotensor. Dimensionless real
parameters $\rho=\{\rho_1,\rho_2,\rho_3\}$ satisfy the conditions
$\mid\rho\mid\leq 1$, $\rho_3^2\leq\mid\rho_1\rho_2\mid$. The scalar
parts  of the noises explicitly written in (\ref{ks}) are in a
standard power form \cite{MHD}. The parameters $ g_0, g'_0, g''_0$
play the role of the bare coupling constants, and $ a,\epsilon$ are
free parameters of the model. The value $\epsilon=2$ corresponds to
the Kolmogorov energy pumping from infrared region of the small
${\bf k}$.

\section{Renormalization}

In a standard way, one can solve the primary infrared problem for
the physical value $\epsilon=2$ by the transfer to the region of
small values $\epsilon$, where the ultraviolet (UV) divergences
appear. They can be eliminated by the addition of the appropriate
counterterms to the action (\ref{zd}) \cite{RG91}. The counterterms
are formed of the superficial UV divergences, which are present in
one-particle irreducible Green functions \cite{ren}. The following
1-PI Green functions possess the  UV divergences: $ <{\bf v}'{\bf
v}>, \, <{\bf b}'{\bf b}>, \, <{\bf v}'{\bf b}>, \, <{\bf b}'{\bf
v}>, \, <{\bf v}'{\bf b}{\bf b}> $.
Corresponding diagrams are shown in Fig.\,1 and Fig.\,2 and
counterterms have the form:
$\nu{\bf v}'\Delta{\bf v}$, $\nu{\bf b}'\Delta{\bf b}$, ${\bf
v}'({\bf b}\partial){\bf b}$, $ \nu{\bf v}'\Delta{\bf b}$ and
$\nu{\bf b}'\Delta{\bf v} $. The  last two of them are not present
in the primary action (\ref{zd}). For this reason, it is necessary
to consider the extended theory with the additional cross
dissipative terms $ v\nu{\bf v}'\Delta{\bf b}$, $ w\nu{\bf
b}'\Delta{\bf v} $ and helical magnetic Prandtl numbers $ v^{-1} $,
$ w^{-1} $. Besides UV divergences  mentioned above which manifest
themselves like the poles of $\epsilon$ ($\epsilon-$UV divergences),
another divergences proportional to the UV cutoff $\Lambda$   can
appear in the Green functions $ <{\bf v}'{\bf v}>,$ $<{\bf v}'{\bf
b}>,$ $ <{\bf b}'{\bf v}>,$ $ <{\bf b}'{\bf b}>$. They acquire the
form of
 $\Lambda{\bf v}'\, rot \,{\bf v}$,
 $ \Lambda{\bf v}'\, rot\,{\bf b}$,
 $ \Lambda{\bf b}'\, rot\,{\bf v}$ and
 $ \Lambda{\bf b}'\, rot\,{\bf b}.$
These $\Lambda-UV$ divergences generate the instability of the
model, which causes exponential growth in time of the corresponding
response functions. Therefore, their direct insertion into the
action (\ref{zd}) is not allowed and one has to find an effective
way to eliminate them. One can make a natural assumption that
finally the energy of the unstable large scale magnetic fluctuations
must be transformed into the energy of large scale mean magnetic
field.

\input epsf
   \begin{figure}[t]
     \vspace{0cm}
       \begin{center}
       \leavevmode
       \epsfxsize=10cm
       \epsffile{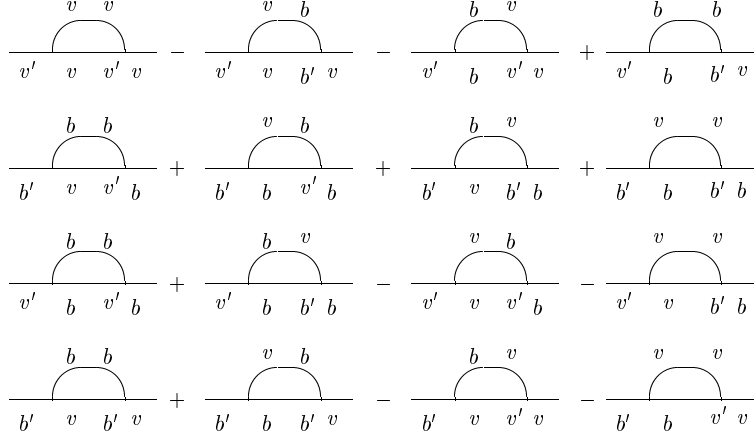}
   \end{center}
\vspace{0cm} \caption{One-loop Feynman diagrams which are
UV-divergent. Only diagrams related to the Green functions $ <{\bf
v}'{\bf v}>,$ and $<{\bf b}'{\bf b}>$  (first and second line) can
contain $\Lambda-UV$ divergences}
\end{figure}


The model of the helical MHD under consideration describes steady
state, therefore it is reasonable to consider the new vacuum state
with zero mean values of fields ${\bf v},$ ${\bf v'},$ ${\bf b'},$
and non-vanishing time-independent mean field $<{\bf b}>\equiv {\bf
c} \ne 0.$

\input epsf
   \begin{figure}[t]
     \vspace{0cm}
       \begin{center}
       \leavevmode
       \epsfxsize=10cm
       \epsffile{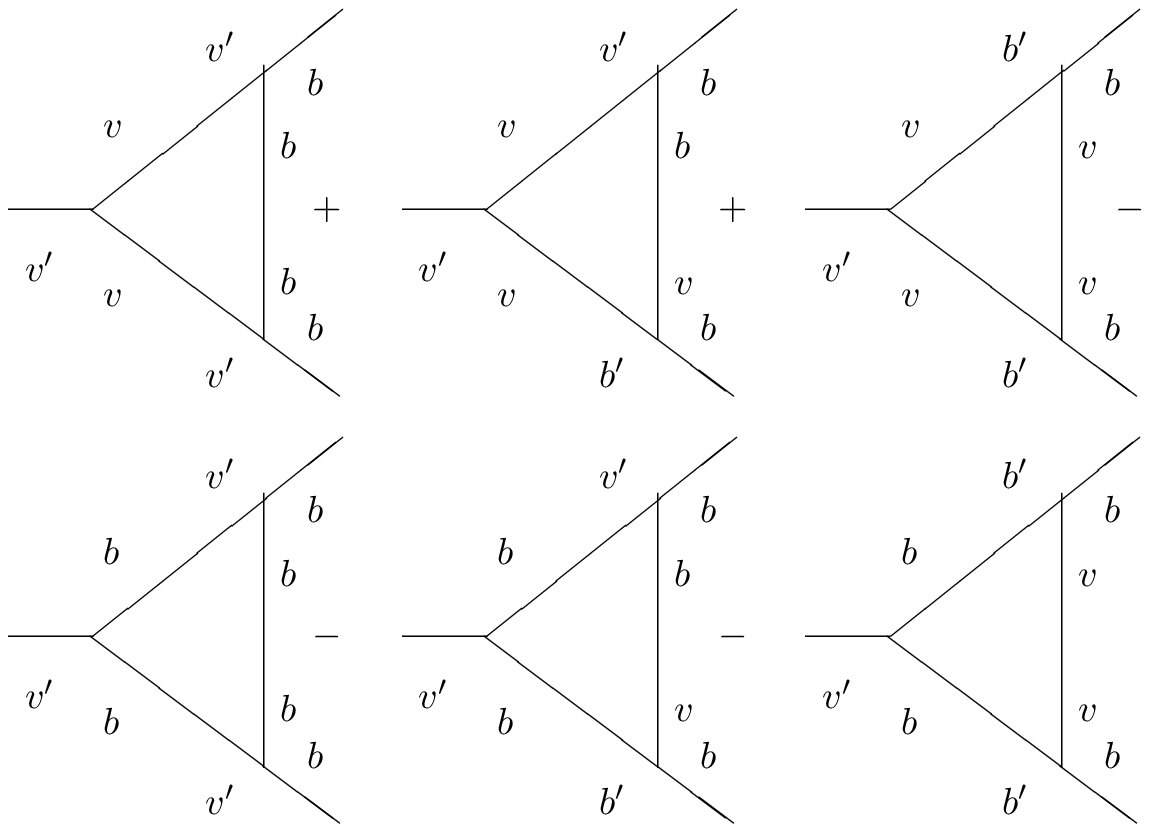}
   \end{center}
\vspace{0cm} \caption{Diagrams related to the Green function $<{\bf
v}'{\bf b}{\bf b}>$.}
\end{figure}


In a quantum field theory the appearance of non-zero vacuum value of
field is associated with spontaneous symmetry breaking
\cite{Vasiliev}, and as a standard, the value itself is determined
from requirement of minimum of potential energy at the tree level.
In the  case considered here the situation is more complicated and
rather technically different, because unstable $\Lambda$-terms
appear at the next (one-loop) level, consequently, the nonvanishing
value of mean magnetic field in the steady state must be calculated
in this order of a perturbation scheme. By straightforward
calculations  and/or from the symmetry analysis of the given
one-loop Feynman diagrams, one can find that only $<{\bf b'}{\bf
b}>$ contains $\Lambda-$terms. Here all $\Lambda-$ divergences can
be eliminated by means of the shift of ${\bf b},$ namely ${\bf
b(x)}\rightarrow {\bf b(x)}+{\bf c}$, and, on the other hand,
$\epsilon-$UV divergences are compensated by means  of five
independent renormalization constants $Z_i, i=1, ...5$ in extended
model of helical MHD. As a result one obtains model with
renormalized action:
\begin{eqnarray}
S^h_{R}(\Phi)&=&\frac{{\bf v^{\prime}} D^{{\bf vv}} {\bf
v^{\prime}}}{2}+ \frac{{\bf b^{\prime}} D^{{\bf bb}} {\bf
b^{\prime}}}{2}+ \frac{{\bf v^{\prime}} D^{{\bf vb}} {\bf
b^{\prime}}}{2}+ \frac{{\bf b^{\prime}} D^{{\bf bv}} {\bf
v^{\prime}}}{2}- {\bf v^{\prime}} [\partial_t {\bf v} -
\nonumber \\
&&  Z_1\nu \triangle {\bf v}- Z_4 v \nu \triangle {\bf b} +({\bf
v}\partial){\bf v}- Z_3({\bf b}
\partial){\bf b}-Z_3 ({\bf c}\partial){\bf b}] -
\label{action1} \\
&& {\bf b^{\prime}}[\partial_t {\bf b}-Z_2 u \nu \triangle {\bf b}-
Z_5 w \nu \triangle {\bf v} +({\bf v}\partial){\bf b}- ({\bf
b}\partial){\bf v}- ({\bf c}\partial){\bf v}], \nonumber
\end{eqnarray}
where all parameters are renormalized couterparts of bare ones. The
action (\ref{action1}) generates Green functions without
divergences. In this case Feynman rules have the following form
\begin{eqnarray}
\Delta^{vv'}_{12}&=&\frac{M P_{12}}{L M-S V},\quad\quad\quad\quad
\Delta^{vb'}_{12}=-\frac{V P_{12}}{L M -S V}\,,\nonumber\\
\Delta^{bv'}_{12}&=&-\frac{S P_{12}}{L M -S V},\quad\quad\quad\quad
\Delta^{bb'}_{12}=\frac{L P_{12}}{L M -S V}\,,\nonumber\\
\Delta^{v'v}_{12}&=&\frac{M^+ P_{12}}{L^+ M^+ -S^+ V^+},\quad\quad
 \Delta^{b'v}_{12}=-\frac{V^+ P_{12}}{L^+ M^+
-S^+ V^+}\,,\nonumber\\
 \Delta^{v'b}_{12}&=&-\frac{S^+ P_{12}}{L^+ M^+ -S^+
V^+},\quad\,\,
 \Delta^{b'b}_{12}=\frac{L^+ P_{12}}{L^+ M^+ -S^+
V^+}\,,\nonumber\\
 \Delta^{vv}_{12}&=&\frac{D^{vv}_{12}M M^+
-D^{vb}_{12}M V^+ -D^{bv}_{12} V M^+ +D^{bb}_{12}V V^+}{(L^+ M^+
-S^+ V^+)(L M-S V)}\,, \nonumber \\
 \Delta^{bb}_{12}&=&\frac{D^{vv}_{12}S S^+
-D^{vb}_{12}S L^+ -D^{bv}_{12} L S^+ +D^{bb}_{12}L L^+}{(L^+ M^+
-S^+ V^+)(L M-S V)}\,, \nonumber \\
 \Delta^{vb}_{12}&=&\frac{-D^{vv}_{12}M S^+
+D^{vb}_{12}M L^+ +D^{bv}_{12} V S^+ -D^{bb}_{12}V L^+}{(L^+ M^+
-S^+ V^+)(L M-S V)}\,, \nonumber \\
 \Delta^{bv}_{12}&=&\frac{-D^{vv}_{12}S M^+
+D^{vb}_{12}S V^+ +D^{bv}_{12} L M^+ -D^{bb}_{12}L V^+}{(L^+ M^+
-S^+ V^+)(L M-S V)}\,,
\end{eqnarray}
where
\begin{eqnarray}
L&=&-i\omega+\nu k^2\,,\,\,\,M=-i\omega+\nu u k^2\,,\nonumber\\
V&=&\nu v k^2 -i \gamma\,,\,\,\,\,\,\,S=\nu w k^2
-i\gamma\,,\nonumber\\ D^{vv}_{mn}&=&g_1 \nu^3
k^{4-d-2\varepsilon}(P_{mn}+i\rho_1
\varepsilon_{mnl}\frac{k_l}{k})\,,\nonumber\\
 D^{vb}_{mn}&=&g_3
\nu^3 k^{4-d-\varepsilon(1+a)}(P_{mn}+i\rho_3
\varepsilon_{mnl}\frac{k_l}{k})\,,\nonumber\\
 D^{bv}_{mn}&=& D^{vb}_{mn}\,,\nonumber\\
 D^{bb}_{mn}&=&g_2
\nu^3 k^{4-d-2\varepsilon a}(P_{mn}+i\rho_2
\varepsilon_{mnl}\frac{k_l}{k})\,,
\end{eqnarray}
with
\begin{equation}
\gamma={\bf c} \cdot {\bf k}\,.
\end{equation}
and vertices are defined by the expressions:
\begin{eqnarray}
{\bf v}'\cdot ({\bf v}\cdot {\bf
\partial}){\bf v}&=&v_i't_{ijl}v_j v_l /2\,,\nonumber \\
 {\bf v}'\cdot ({\bf b}\cdot {\bf \partial}){\bf b}&=&v_i't_{ijl}b_j b_l
 /2\,,\nonumber \\
{\bf b}'\cdot({\bf b}\cdot {\bf \partial}) {\bf v}- {\bf
b}'\cdot({\bf v}\cdot {\bf \partial}) {\bf b}&=& b_i'\bar{t}_{ijl}
b_j v_l\,,
\end{eqnarray}
where
\begin{equation}
t^k_{ijl}=i(k_j\delta_{il}+k_l\delta_{ij})\,,\quad
\bar{t}^k_{ijl}=i(k_j\delta_{il}-k_l\delta_{ij})\,.
\end{equation}
Using Feynman rules defined above one can immediately calculate the
diagrams which are shown in Fig.\,1 and Fig.\,2. To extract the
$\Lambda-$ terms it is enough to keep only linear in wave vector
part of the diagrams. The response functions $<{\bf v'} {\bf v}>$,
$<{\bf v'} {\bf b}>$, $<{\bf b'} {\bf v}>$ do not possess any linear
divergent terms. $\Lambda$-divergent part and term $\sim {\bf c}$ of
the response function $<{\bf b'} {\bf b}>$ at the one-loop level are
of the form:
\begin{eqnarray}
<b'_i b_j> &\sim& i  k_m \varepsilon_{iml}(g \rho_1 C_1+g' \rho_2
C_2+g'' \rho_3 C_3) \times \nonumber \\ &\times& \left[\nu \Lambda
\delta_{jl}-|{\bf c}|\frac{3\pi}{8}
\sqrt{\frac{(1+u)^2-(v+w)^2}{(1+u)^2(u-v w)}}
\left(\delta_{jl}+e_je_l\right)\right], \label{sigma}
\end{eqnarray}
where ${\bf e}\equiv {\bf c}/|{\bf c}|,$ $C_1=(w(v+w)-u(1+u))/\xi,$
$C_2=(1+u-v(v+w))/\xi,$ $C_3=2(u v-w)/\xi$ and $\xi=6(1+u)^2(u-v
w)\pi^2$. From  requirement of vanishing of $\Lambda-$term in
(\ref{sigma}) one determines  the value of spontaneous field
\begin{equation}
|{\bf c}|=\frac{8\nu}{3\pi} \sqrt{\frac{(1+u)^2-(v+w)^2}{(1+u)^2(u-v
w)}}\Lambda\,. \label{pole}
\end{equation}
One can see from this equation that magnitude of spontaneous field
is independent of coupling constants
$g,g^{\prime},g^{\prime\prime}$, which characterize an intensity of
random noises, and helical parameters $\rho_1,\rho_2,\rho_3$. In
such a way we obtain the renormalized Green functions which are
finite as $\Lambda \rightarrow \infty$ formally, as it is usual in
the field theory. But in real problems a natural maximal cutoff
exists. In the developed turbulence the Kolmogorov dissipative
length $l_D=\Lambda^{-1}$ plays the role of a minimal scale. This
length can be expressed in terms of basic phenomenological
parameters - viscosity $\nu$ and energy dissipation rate
$\varepsilon$; $l_d=\nu^{3/4}\varepsilon^{-1/4}$. Then from
(\ref{pole}) one obtains $|{\bf c}| \sim (\nu \varepsilon)^{1/4}$
and it determines the order of magnitude of the spontaneous field
${\bf c}$.

\section{Corrections to the Alfv\'en waves}

In order to understand the role of the last term in (\ref{sigma})
let us consider the linearized  MHD equations which follow from
(\ref{action1}) in infrared limit of small wave vector ${\bf k}$:
\begin{eqnarray}
\partial_t {\bf v}&=&-\nu k^2 {\bf v}+
(v \nu k^2 +i\gamma){\bf b}
\nonumber \\
\partial_t {\bf b}&=& (-w \nu k^2+i\gamma){\bf v}
-u \nu k^2{\bf b} + i\chi [{\bf k}\times {\bf e}] ({\bf b} \cdot
{\bf e})  \, . \label{exot}
\end{eqnarray}
Here $\chi \equiv 3(g \rho_1 C_1+g' \rho_2 C_2+g'' \rho_3 C_3) |{\bf
c}|\sqrt{((1+u)^2-(v+w)^2)/(u-v w)}/$ $(8\pi(1+u))$ and
$\gamma\equiv ({\bf k}{\bf c}).$ To solve Eq.\,(\ref{exot}) we
define a basis of orthonormal vectors ${\bf n}\equiv {\bf k}/k,$ $
{\bf l}\equiv ({\bf e}-{\bf n} \cos \delta)/\sin\delta,$ ${\bf
m}\equiv [{\bf n}\times {\bf e}]\sin\delta,$  where $\delta$ is the
angle between vectors ${\bf n}$ and ${\bf e}.$ The transversal
fields ${\bf v}, {\bf b}$ can be decomposed with respect to the
vectors ${\bf v}, {\bf b}$: ${\bf v}= v_l {\bf l}+v_m{\bf m},$ ${\bf
b}=b_l{\bf l}+ b_m{\bf m}.$ Time dependent amplitudes $v_l, v_m,
b_l, b_m$ satisfy equation (\ref{exot}) and are given by the
following expressions:
\begin{eqnarray}
v_l&=&\frac{1}{c}\left[c_1 e^{\lambda_2 t}(\lambda_2-d)+c_2
e^{\lambda_1 t}(\lambda_1-d)\right]\,, \\
 b_l&=&c_1 e^{\lambda_2 t}+c_2 e^{\lambda_1 t}\,,\\
 v_m&=&\frac{1}{2 \lambda_1 -a-d}\nonumber \\
 &\times&\Biggl(e^{\lambda_2 t} \left(-c_1 b e
 \left(\frac{1}{2 \lambda_1 -a-d}+t\right)+
 c_3 \left(\lambda_1-a\right)-c_4 b\right) + \nonumber \\
 &+& e^{\lambda_1 t} \left(-c_2 b e
 \left(\frac{1}{2 \lambda_1 -a-d}-t\right)+
 c_3 \left(\lambda_1-d\right)+c_4 b\right)\Biggl)\,,\\
b_m&=&\frac{1}{2 \lambda_1 -a-d}\Biggl(e^{\lambda_2 t} \Biggl(c_1
\frac{e}{2}
 \left(\frac{a-d}{2 \lambda_1 -a-d}-1+
 2 t\left(\lambda_1-d\right)\right) \nonumber \\
 &-&c_3 c+c_4\left(\lambda_1-d\right)\Biggr)+
 e^{\lambda_1 t} \Biggl(c_2 \frac{e}{2}
 \left(\frac{a-d}{2 \lambda_1 -a-d}+1+
 2 t\left(\lambda_1-a\right)\right) \nonumber \\
 &+&c_3 c+c_4\left(\lambda_1-a\right)\Biggr)
\Biggl)\,, \label{hura}
\end{eqnarray}
where $c_1, c_2, c_3$ and $c_4$ are constants of integration and we
have defined
\begin{eqnarray}
\lambda_{1,2}&=&\frac{1}{2}(a+d\pm\sqrt{(a+d)^2-4(a d-b c)})\,,\\
a&=&-\nu k^2\,,\nonumber \\ b&=&i \gamma -v \nu k^2\,, \nonumber
\\ c&=&i \gamma -w\nu k^2\,,\nonumber \\ d&=&-u \nu k^2\,,
\nonumber
\\
e&=&-i k \sin^2\delta \frac{3}{4 (2 \pi)^2}|c|(g_1 \rho_1
C_1+g_2\rho_2 C_2+g_3 \rho_3 C_3)\frac{K}{\xi_1 (1+u)} \,,\nonumber
\end{eqnarray}
where
\begin{eqnarray}
K&=&\sqrt{\frac{(1+u)^2-(v+w)^2}{u-v w}}\,, \nonumber \\
 \xi_1&=&(1+u)^2(u-v w)\,.\nonumber
\end{eqnarray}

The functions $\lambda_{1,2}$ are complex with negative real parts
for the physical values of Prandtl numbers $u^{-1}, v^{-1}, w^{-1},$
and they suppress the linear increase of amplitudes. Therefore, the
last term in (\ref{sigma}) results in the appearance of specific
long-lived pulses of Alfv\'en waves. They are orthogonal  polarized
with respect to the spontaneous field and try to restore the
isotropy broken by them.

\section{Short remarks to the RG analysis and critical regimes}

After  successful elimination of all UV divergences  we can use the
renormalization group procedure and arrive to the set of RG
Gell-Mann-Low equations for five invariant charges:
\begin{eqnarray}
\label{gle} s\frac{d\bar g_i(s)}{ds} & = & \beta_{g_i}(\bar g(s),
\epsilon)\,, \qquad \bar g_i(s)\mid_{s=1}=g_i\,\, g_i\equiv g, g',
u, v, w\,,
\end{eqnarray}
where $ s=p/m$ ($m$ is scale setting parameter). RG
$\beta$-functions are expressed via renormalization constants $Z_i,$
which have been calculated in one-loop approximation. Note that the
$\beta$-functions are finite in the limit of $\epsilon\to 0.$

The Gell-Mann-Low equations (\ref{gle}) have been solved numerically
for the various initial values of the invariant charges $g_i$. It
provides the possibility to analyze the attracting regions of
infrared fixed points. There are two infrared-stable fixed points:
the Gaussian $ g_i^*=0$ and nontrivial $ g^*\ne0$, $ u^*=1.393$, $
g'^*=u^*=v^*=w^*=0$. The latter  provides the existence of
asymptotic critical regime of the Kolmogorov type. The RG approach
improves expressions of simple perturbation theory and leads to the
replacement of the original charges by the invariant ones, and, in
critical regime by their values at the corresponding fixed points.
Keeping in mind that ${\bf c}$ is connected with magnetic induction
${\bf B}$ by the relation ${\bf c}={\bf B}/\sqrt{4 \pi \rho \mu}$
(remember that $\rho$ denotes density and $\mu$-permeability of
fluid) in above asymptotic region the magnitude of mean spontaneous
magnetic induction is equal to $\frac{16}{3}\sqrt{\frac{\rho
\mu}{\pi u_*}}(\nu \varepsilon)^{1/4}$.

\section{Conclusion}

In the paper the correlation and response functions of velocity and
magnetic fluctuations have been studied. Generally, these functions
possess singularities, which   can   be   eliminated   by a proper
renormalization procedure. As a result, RG equations have been
obtained and their solution have been found in the range of small
wave numbers. This solution corresponds to the famous Kolmogorov
scaling law. In helical MHD, where mirror symmetry of the system
under consideration is stochastically broken, the non-vanishing mean
magnetic field is spontaneously generated. This phenomena is
accompanied by the linear growth in time $t$ of the amplitudes of
Alfv\'en waves for small wave vectors. Due to the viscosity terms
this growth is transformed into long-lived pulses of the type $t
\exp(-i \beta t)\exp (-\alpha t)$ with small $\alpha>0$ and $\beta$.

\vspace{0.8cm} \noindent {\bf Acknowledgments} \vspace{0.3cm}

\noindent This work was supported by Slovak Academy of Sciences
within the project 7232. M.H. gratefully acknowledges the
hospitality of the N.N.Bogoliubov Laboratory of Theoretical Physics
at JINR Dubna.


\begin{thebibliography}{99}
\bibitem{vas} Adzhemyan, L.Ts., Antonov, N.V., Vasil'ev, A.N. (1999)
 {\it The Field Theoretic Renormalization Group in Fully Developed
 Turbulence}, (Gordon and Breach Sci. Publ., The Netherlands).
\bibitem{MHD} Adzhemyan, L.Ts., Vasil'ev, A.N., Hnatich, M. (1985)
 Quantum-field renormalization group in the theory of turbulence:
 Magnetohydrodynamics,
 {\it Teor. Mat. Fiz.}, {\bf 64}, pp.196-207
\bibitem{four} Fournier, J.D., Sulem, P.L., Pouquet, A. (1982)
Infrared properties of forced magnetohydrodynamic turbulence, {\it
J. Phys.} Math. Gen., {\bf A 15}, pp.1393-1420.
\bibitem{RG91} Hnatich, M.,  Stehlik, M. (1992)
Renormalization  group in gyrotropic magnetic  hydrodynamics,  In
"Renormalization  group '91".  Eds. Shirkov D.V., Priezzev V.B.,
World Scien. Pub.,  Singapore, 204.
\bibitem{tur} Vajnshtein, S.I., Zeldovich, Ja.B., Ruzmajkin, A.A. (1980)
 {\it Turbulent Dynamo in Astrophysics}, (Nauka, Moscow).
\bibitem{dyn} Adzhemyan, L.Ts., Vasil'ev, A.N., Hnatich, M. (1987)
 Turbulent dynamo as spontaneous symmetry breaking,
 {\it Teor. Mat. Fiz.}, {\bf 72}, pp.369-383
\bibitem{MSR1} De Dominicis, C., Martin, P.C. (1979)
 Energy spectra of certain randomly-stirred fluids,
 {\it Phys. Rev.}, {\bf A 19}, pp.419-422
\bibitem{MSR} Adzhemyan, L.Ts., Vasil'ev, A.N., Pis'mak, Yu.M. (1983)
 Renormalization-group approach in the theory of turbulence:
 The dimensions of composite operators,
 {\it Teor. Mat. Fiz}, {\bf 57}, pp.268-281
\bibitem{ren} Collins, J.C. (1984) {\it Renormalization}.
 Cambridge University press, London.
\bibitem{Vasiliev}
Vasiliev, A.N. (1998) {\it Quantum field renormalization group in
theory of critical phenomena and stochastic dynamics}, (Sankt
Peterburg),(in Russian).

\end{thebibliography}
\end{document}